\documentclass[english]{article}
\usepackage[T1]{fontenc}
\usepackage[latin1]{inputenc}
\usepackage{graphics}

\makeatletter

\providecommand{\LyX}{L\kern-.1667em\lower.25em\hbox{Y}\kern-.125emX\@}

 \newcommand{\lyxaddress}[1]{
   \par {\raggedright #1 
   \vspace{1.4em}
   \noindent\par}
 }

\usepackage[T1]{fontenc}
\usepackage[latin1]{inputenc}
\usepackage{graphics}

\makeatletter

\usepackage{graphics}

\makeatletter

\usepackage[T1]{fontenc}
\usepackage{babel}
\usepackage{graphics}

\makeatletter


\makeatother

\makeatother
\makeatother

\makeatother
\begin{document}
\setcounter{footnote}{0}\begin{titlepage}

\title{Sphaleron in the Dilatonic Gauge Field Theory}

\author{D.Karczewska\thanks{
dkarcz@us.edu.pl 
}, R.Manka\thanks{
manka@us.edu.pl 
}}

\maketitle

\lyxaddress{Department of Astrophysics and Cosmology, \\
 Institute of Physics, \\
 University of Silesia, \\
 Uniwersytecka 4, 40-007 Katowice,\\
 Poland}

\begin{abstract}
Motivated by the Kaluza-Klein theory with a large number of extra
spacetime dimensions, we present a numerical study of static, spherically
symmetric sphaleron solutions coupled to the dilaton fields. We show
that sphalerons may have different dilaton cloud configurations, resulting
in new massive sphalerons, in general. However, there exist different
cloud configurations with different values of the dilaton mass. 
\end{abstract}
\vskip 4cm 

~~~~~~

PAC numbers(s): 98.80.Cq,12.15.Cc 

\end{titlepage}

\renewcommand{\thepage}{\arabic{page}}

\setcounter{page}{1} \renewcommand{\thefootnote}{\arabic{footnote}} \setcounter{footnote}{0}

\section{INTRODUCTION}

Recently there has been a considerable interest in the field theories
with large number of extra spacetime dimensions. In comparison to
the standard Kaluza-Klein theory these extra dimensions may be restricted
only to the gravity sector of the theory while the Standard Model
(SM) fields are assumed to be localized on the 4-dimensional spacetime
{[}1-3{]}. It is a promising scenario from the phenomenological point
of view because it shifts the energy scale of unification from \( 10^{19}\, GeV \)
down to \( 10-100\, TeV \). 

The gauge field theory is extended by inclusion of the dilatonic field
in such theories. Such fields appear also in a natural way in the
Kaluza-Klein theories \cite{tap:1987}, superstring inspired theories
\cite{ew:1985,sf:1989} and in theories based on the non-commutative
geometry approach \cite{abc:1993}. 

As previous studies have already shown the inclusion of dilatons in
a pure Yang - Mills theory has consequences already at the classical
level. In particular, the dilaton Yang - Mills theories possess 'particle
- like' solutions with finite energy which are absent in pure Yang
- Mills case. 

On the other hand, the sphaleron was introduced by Klinkhamer and
Manton \cite{km:1983} to describe a static electroweak gauge - field
configuration that constitutes a saddle point between two vacua, differing
from one another by non trivial topology (the hedgehog topology). 

Analogous equations have recently been obtained for the 't Hooft -
Polyakov monopole model coupled to the dilatonic field \cite{fg:1996}.
There is also growing interest \cite{krs:1985} in baryon number violation
within the Standard Model induced by sphalerons. The rate of baryon
number violating processes depends on the energy of the sphaleron
\cite{am:1988}. 

The aim of this paper is to examine the properties of the sphaleron
solution in the presence of a dilatonic field, in the electroweak
theory. We shall demonstrate below the existence of spherically symmetric
dilatonic clouds surrounding the sphalerons with interesting implications.

\section{THE DILATONIC GAUGE FIELD THEORY}

Dilatons appear in the higher dimensional theory after the process
of compactification. To illustrate the point, we consider six-dimensional
Kaluza - Klein theory. 

Let us now consider the action integral of Einstein-Yang-Mills-Higgs
theory in the six-dimensional spacetime: \begin{equation}
\label{a1}
{\mathcal{S}}=\int d^{6}x\sqrt{-g_{6}}L,
\end{equation}
 where \( g_{6}=det(g_{MN}) \) and \( M=\{\mu ,\, i\} \), \( N=\{\nu ,\, j\} \)
with \( x^{M}=\{x^{\mu },y^{i}\},\, i=1,2 \). The metrical tensor
in the six-dimensional spacetime can be written: \begin{equation}
\label{a2}
g_{MN}=\left( \begin{array}{cc}
e^{-2\xi (x)/f_{0}}\overline{g}_{\mu \nu } & 0\\
0 & -\delta _{ij}e^{+2\xi (x)/f_{0}}
\end{array}\right) ,
\end{equation}
 where \( f_{0} \) is an arbitrary constant. According to the above
definition we can write: \begin{equation}
\label{a3}
\sqrt{-g_{6}}=\sqrt{-\overline{g}}e^{-2\xi (x)/f_{0}}.
\end{equation}
 In equation (\ref{a2}) \begin{equation}
g_{\mu \nu }=e^{-2\xi (x)/f_{0}}\overline{g}_{\mu \nu }
\end{equation}
 represents the four-dimensional metric in the Jordan frame, while
\( \overline{g}_{\mu \nu } \) refers the Einstein frame. Fluctuations
around the four-dimensional Minkowski \begin{equation}
\overline{g}_{\mu \nu }=\eta _{\mu \nu }+h_{\mu \nu }(x,y)
\end{equation}
 will produce the interaction with Kaluza-Klein dilatons \begin{equation}
\label{grav}
h_{\mu \nu }(x,y)=\sum _{{\textbf {n}}}h_{\mu \nu }^{{\textbf {n}}}(x)e^{i\frac{2\pi n^{i}y^{i}}{r_{2}}}
\end{equation}
 with the typical mass scale \( M \) (for \( n^{i}\neq 0 \)). 

We consider here the Lagrangian of the Einstein-Yang-Mills-Higgs field
as follows: 

\begin{eqnarray}
 & L=L_{g}+L_{SM}\delta (y), & \\
 & L_{g}=-\frac{1}{2\kappa _{6}}R, & \\
 & L_{SM}=-\frac{1}{4}F_{MN}^{a}F^{aMN}+D_{M}H^{+}D^{M}H-U(H), & 
\end{eqnarray}
 where \( \kappa _{6} \) is the six-dimensional gravitional coupling
and \begin{equation}
U(H)=\lambda \left( H^{+}H-\frac{1}{2}v_{0}^{2}\right) ^{2}.
\end{equation}
 \( R \) is a curvature scalar defined as usual: \begin{equation}
\label{a5}
R=g^{MN}R_{MN},
\end{equation}
 and \( R_{MN} \) is the Ricci tensor: \begin{equation}
\label{a6}
R_{MN}=\partial _{L}\Gamma _{MN}^{L}-\partial _{N}\Gamma _{ML}^{L}+\Gamma _{MN}^{L}\Gamma _{LR}^{R}-\Gamma _{ML}^{R}\Gamma _{NR}^{L}.
\end{equation}
 \( \Gamma _{MN}^{L} \) - are the six-dimensional Christoffel symbols.
Let us compactify the six-dimensional spacetime to the four-dimensional
Minkowski one on the torus \( ({\mathcal{M}}_{6}\rightarrow {\mathcal{M}}_{4}\times {\mathcal{S}}^{1}\times {\mathcal{S}}^{1}) \).
In this paper we assume that the extra dimensions are compactified
to a two-dimensional torus with a single radius \( r_{2} \). For
the four-dimensional Minkowski spacetime (\( \overline{g}_{\mu \nu }=\eta _{\mu \nu } \))
\begin{equation}
R=\frac{4}{f_{0}^{2}}e^{2\xi (x)/f_{0}}\{-\partial _{\mu }\xi \partial ^{\mu }\xi +f_{0}\partial _{\mu }\partial ^{\mu }\xi \}.
\end{equation}
 The six-dimensional action may be rewritten as: \begin{equation}
\label{a7}
{\mathcal{S}}=\int d^{4}x\int d^{2}y\sqrt{-g_{6}}L=\int d^{4}x\sqrt{-\overline{g}}{\mathcal{L}},
\end{equation}
 where \( \int d^{2}y=(2\pi r_{2})^{2} \) and \( {\mathcal{L}} \)
is the effective Lagrange function in four-dimensional spacetime.
The six-dimensional gravitional coupling \( \kappa _{6}=8\pi G_{6} \)
is convenient to define as \[
G_{6}^{-1}=\frac{4\pi }{(2\pi )^{2}}M^{4},\]
 where \( M \) is the energy scale of the compactification (\( \sim 10-100\, TeV \)).
The cosmological consideration \cite{hall} gives the bound \( M>100\, TeV \)
which corresponds to \( r_{2}<5.1\, 10^{-5}\, mm \). By denoting
the four-dimensional coupling constant \( \kappa =8\pi G_{N}=8\pi M^{-2}_{Pl} \),
we get \begin{equation}
M^{2}_{Pl}=4\pi M^{4}r_{2}^{2}.
\end{equation}
 The parameter \( f_{0} \) is determinated by the Planck mass: \begin{equation}
f^{2}_{0}=2M^{4}r^{2}_{2}=\frac{1}{2\pi }M^{2}_{Pl}
\end{equation}
 so as to produce the \( 1/2 \) term for the dilaton field in (\ref{a9}).
At present we have \begin{equation}
\label{fo}
f_{0}=\frac{1}{\sqrt{2\pi }}M_{Pl}\sim 4.87\, 10^{18}\, GeV/c^{2},
\end{equation}
 but here the Planck mass \( M_{Pl} \) is no longer a fundamental
constant, it may change during the evolution of the universe \cite{eef}.
Cosmological considerations \cite{eef} suggest that in the early
universe the internal radius \( r_{2} \) was smaller (\( r_{2}(t)=b(t)r_{2} \)
where \( b(t) \) is the radion field). As a result, the Planck mass
\( M_{Pl} \) was smaller in the early universe, and the gravitational
interaction bigger than at the present time. An estimation \cite{eef}
of the Planck mass in the early universe gives \[
M_{Pl,0}\approx 10^{-11}\, M_{Pl}\sim 10^{8}\, GeV.\]
 This also leads to a lowering of the \( f_{0} \). 

As a result of compactification of the six-dimensional Lagrangian,
we get the following Lagrangian for the Yang-Mills-Higgs fields :
\begin{equation}
\label{a8}
{\mathcal{L}}_{SM}=(D_{\mu }H)^{+}D^{\mu }H-e^{-\frac{2\xi (x)}{f_{0}}}U(H)-\frac{1}{4}e^{\frac{2\xi (x)}{f_{0}}}F_{\mu \nu }^{a}F^{a\mu \nu }+..
\end{equation}
 and \begin{equation}
\label{a9}
{\mathcal{L}}_{g}=\frac{1}{2}\partial _{\mu }\xi (x)\partial ^{\mu }\xi (x)+...
\end{equation}
 for gravity. 

In this paper we shall focus on the electroweak theory with the gauge
field extended by adding the dilatonic field. One of the interesting
features of the Standard Model, with \( SU_{L}(2)\times U_{Y}(1) \)
symmetry, is the classical scale invariance of the highly symmetric
phase. Then, the anomalous symmetry and quantum effects cause its
break up, and produce nonvanishing cosmological constant. The classical
scale invariance offers a link between the Standard Model and gravity.
The Jordan-Brans-Dicke theory \cite{jbs:1959} of the scalar-tensor
theory of gravity is an example of successful implementation of this
idea. 

The Glashow-Weinberg-Salam dilatonic model with \( SU_{L}(2)\times U_{Y}(1) \)
symmetry is described by the Lagrangian \begin{eqnarray}
{\mathcal{L}}_{B} & = & -\frac{1}{4}e^{2\xi (x)/f_{0}}F_{\mu \nu }^{a}F^{a\mu \nu }-\frac{1}{4}e^{2\xi (x)/f_{0}}B_{\mu \nu }B^{\mu \nu }+\label{a10} \\
 &  & \frac{1}{2}{\partial }_{\mu }\xi (x){\partial }^{\mu }\xi (x)+(D_{\mu }H)^{+}D^{\mu }H-U(H)e^{-2\xi (x)/f_{0}}\nonumber 
\end{eqnarray}
 with the \( SU_{L}(2) \) field strength tensor \( F_{\mu \nu }^{a}=\partial _{\mu }W_{\nu }^{a}-\partial _{\nu }W_{\mu }^{a}+g\epsilon _{abc}W_{\mu }^{b}W^{c\nu } \)
and the \( U_{Y}(1) \) field tensor \( B_{\mu \nu }=\partial _{\mu }B_{\nu }-\partial _{\nu }B_{\mu } \).
Dilaton fields increase the participation of the gauge fields in the
Standard Model and change the Higgs potential into: \begin{equation}
\label{a11}
-\frac{1}{4}F_{\mu \nu }^{a}F^{a\mu \nu }\rightarrow -\frac{1}{4}e^{2\xi (x)/f_{0}}F_{\mu \nu }^{a}F^{a\mu \nu },
\end{equation}

\begin{equation}
\label{a12}
U(H)\rightarrow U(H)e^{-2\xi (x)/f_{0}}.
\end{equation}
 The covariant derivative is given by \( D_{\mu }=\partial _{\mu }-\frac{1}{2}igW_{\mu }^{a}\sigma ^{a}-\frac{1}{2}g^{^{\prime }}YB_{\mu } \),
where \( B_{\mu } \) and \( W_{\mu }=\frac{1}{2}W_{\mu }^{a}\sigma ^{a} \)
are local gauge fields associated with \( U_{Y}(1) \) and \( SU_{L}(2) \)
symmetry groups, respectively. \( Y \) denotes the hypercharge, as
before. The gauge group is a simple product of \( U(1)_{Y} \) and
\( SU(2)_{L} \), hence we have two gauge couplings \( g \) and \( g^{\prime } \).
The generators of gauge groups are represented by the unit matrix
for \( U_{Y}(1) \) and Pauli matrices for \( SU_{L}(2) \). In the
simplest version of the Standard Model a doublet of Higgs fields is
introduced as \( H=\left( \begin{array}{c}
H^{+}\\
H^{0}
\end{array}\right)  \), with the Higgs potential: \begin{equation}
\label{a13}
U(H)=\lambda \left( H^{+}H-\frac{1}{2}v^{2}\right) ^{2}e^{-2\xi (x)/f_{0}}.
\end{equation}
 The dilaton fields change the scale of the interacting constant \( \lambda  \),
thus the mass of Higgs bosons is changed; but the scale of the spontaneous
breaking \( v \) is not altered. The \( f_{0} \) parameter in the
Lagrange function (\ref{a10}) determines the dilaton scale. At the
present time \( f_{0} \) is rather high (\ref{fo}), so the interaction
with dilatons can be neglected. However, in the early universe when
the Planck mass \( M_{Pl} \) was smaller (for details see \cite{eef})
the value of the \( f_{0} \) was smaller, as well. For that reason
we choose the intermediate scale \( f_{0}=10^{7}\, \, GeV \), and
the electroweak symmetry breaking scale \( v=246\, \, GeV \). The
form of the potential (\ref{a13}) leads to a vacuum degeneracy, leading
to the presence of the nonvanishing vacuum expectation value of the
Higgs field, and consequently to the presence of fermion and boson
masses. In the spontaneous symmetry breaking process, the Higgs field
also acquires nonzero mass. 

The Euler-Lagrange equations for the Lagrangian (\ref{a10}) are scale
- invariant: \begin{equation}
\label{t1}
x^{\mu }\rightarrow x^{\prime }{}^{\mu }=e^{\frac{u}{f_{0}}}x^{\mu },
\end{equation}

\begin{equation}
\label{t2}
\xi \rightarrow \xi ^{\prime }=\xi +u,
\end{equation}

\begin{equation}
\label{t3}
H\rightarrow H^{\prime }=H,
\end{equation}

\begin{equation}
\label{t4}
W_{\mu }^{a}\rightarrow W_{\mu }^{,a}=e^{-\frac{u}{f_{0}}}W_{\mu }^{a},
\end{equation}

\begin{equation}
\label{t5}
B_{\mu }\rightarrow B_{\mu }^{\prime }=e^{-\frac{u}{f_{0}}}B_{\mu }.
\end{equation}
 These transformations change the Lagrange function in the following
way: \begin{equation}
\label{t6}
{\mathcal{L}}\rightarrow {\mathcal{L}}^{\prime }=e^{-\frac{2u}{f_{0}}}{\mathcal{L}}.
\end{equation}
 This symmetry can be formulated equivalently as a scaling symmetry
on the coordinates, and the dilaton is often denoted as a Goldstone
boson for dilatation. The origin of the symmetry of the equations
of motion is easily understood from the Kaluza--Klein origin of the
action. The scale transformations are equivalent to a rescaling of
the internal dimensions.

\section{THE DILATONIC SPHALERON}

\noindent The dilatonic solutions in the gravity and the gauge field
theory are a subject of growing interest \cite{fg:1996}. Let us now
consider the sphaleron type solution in the electroweak theory with
dilatons. The sphaleron may be interpreted as inhomogeneous spherical
topological solutions of the motion equations in the Standard Model.
We can assume, for simplicity, that \( g'=0 \) \cite{kb:1991}. 

\noindent Now, let us make the ansatz for the sphaleron Higgs field: 

\begin{equation}
H=\frac{1}{\sqrt{2}}vU(x)h(r)\left( \begin{array}{c}
0\\
1
\end{array}\right) ,
\end{equation}
 where \( U(x)=i\sum \sigma ^{a}n^{a} \) and \( n^{a}=\frac{r^{a}}{r} \)
describe the 'hedgehog' structure. This produces a nontrivial topological
charge of the sphaleron. The topological charge is equal to the Chern-Simons
number. Such a 'hedgehog' structure determines the asymptotic shape
of the sphaleron with gauge fields different from zero: \begin{equation}
W_{i}^{a}=\epsilon _{aij}n^{j}\frac{1-f(r)}{gr},\, \, \, \, W_{0}^{a}=0.
\end{equation}
 We can define the dilatonic field: \begin{equation}
\xi (x)=f_{0}s(x).
\end{equation}
 Let us introduce the dimensionless variable \( x \), defined as
\( x=gvr=2M_{W}r={2r}/{r_{W}} \), where \( M_{W}=\frac{1}{2}gv\sim 80\, \, \, \, GeV \),
\( r_{W}=\frac{1}{M_{W}}\sim 10^{-3}fm \). The spherical symmetry
is assumed for the dilaton field \( s(x) \) as well as for the Higgs
field \( h(x) \) and the gauge field \( f(x) \), resulting in the
following expression for the total energy: \begin{equation}
E=\frac{4\pi v}{g}\int \rho _{0}(x)x^{2}dx,
\end{equation}
 where the energy density is: \begin{eqnarray}
\rho _{0}(x) & = & \frac{1}{2}h'(x)^{2}+\frac{1}{2}\beta s'(x)^{2}+\frac{1}{x^{2}}e^{2s(x)}\left\{ f'(x)^{2}+\frac{1}{2x^{2}}(f(x)-1)^{2}(f(x)-3)^{2}\right\} \nonumber \\
 & + & \frac{1}{4}\varepsilon \left( h(x)^{2}-1\right) ^{2}e^{-2s(x)}+\frac{1}{4x^{2}}(f(x)-3)^{2}h(x)^{2}.
\end{eqnarray}
 \( M_{H}^{2}=2\lambda v^{2} \) determines the Higgs mass; \[
\beta =\frac{f_{0}^{2}}{v^{2}}=\frac{2M^{4}r^{2}_{2}}{v^{2}}\sim 10^{9}\]
 and \( \varepsilon =\frac{M_{H}^{2}}{2M_{W}^{2}}\sim \, \, 1.152 \)
(for the Higgs mass \( M_{H}\sim 120 \) GeV ) is a dimensionless
parameter which determines the sphaleron system completely. As a result,
the Euler-Lagrange equations are: 

\begin{eqnarray}
f^{\prime \prime }(x) & + & 2f'(x)s'(x)+\frac{1}{4}(3-f(x))h(x)^{2}e^{-2s(x)}\\
 & - & \frac{1}{x^{2}}(f(x)-1)(f(x)-2)(f(x)-3)=0;\nonumber \label{rw1} 
\end{eqnarray}
 the \( f(x) \) function describes the gauge field inside the sphaleron,
and the \( h(x) \) function describes the Higgs field in our theory.
The last one satisfies the following equation: \begin{eqnarray}
h^{\prime \prime }(x)+\frac{2}{x}h'(x)+\varepsilon e^{-2s(x)}(1-h^{2}(x))h(x)-\frac{1}{2x^{2}}(3-f(x))^{2}h(x)=0.\label{rw2} 
\end{eqnarray}
 The \( s(x) \) function describes the dependence of the dilaton
field on \( x \) in the extended electroweak theory and obeys the
equation: \begin{eqnarray}
s^{\prime \prime }(x) & +\frac{2}{x}s^{\prime }(x)+\frac{e^{2s(x)}}{\beta }\left\{ -\frac{2}{x^{2}}f'(x)^{2}-\frac{1}{x^{4}}(f(x)-1)^{2}(f(x)-3)^{2}\right\}  & \\
 & +\frac{\varepsilon }{2\beta }e^{-2s(x)}(1-h^{2}(x))^{2}=0.\nonumber 
\end{eqnarray}
 The simplest solutions are global and correspond to the vacuum with
the broken symmetry in the Standard Model. It is obvious that far
from the center of the sphaleron our solutions should describe the
broken phase which, is very well known from the Standard Model. We
can find easily the asymptotic solutions for \( x\rightarrow 0 \):
\begin{equation}
f(x)=1+2rx^{2}+O(x^{3}),
\end{equation}

\begin{equation}
h(x)=ux+O(x^{3}),
\end{equation}

\begin{equation}
s(x)=(a+s_{\infty }-awx^{2})+O(x^{3}),
\end{equation}
 and for \( x\rightarrow \infty  \): \begin{equation}
f(x)=3-f_{\infty }e^{-x},
\end{equation}

\begin{equation}
h(x)=1-\frac{h_{\infty }}{x}e^{-\sqrt{2\epsilon }},
\end{equation}

\begin{equation}
s(x)=s_{\infty }-\frac{d_{\infty }}{x}.
\end{equation}
 We make an ansatz for the trial functions as: \begin{equation}
f(x)=1+2\tanh ^{2}(\sqrt{r}x)\sim 1+2rx^{2}-\frac{4r^{2}x^{4}}{3}+O(x^{5}),
\end{equation}

\begin{equation}
h(x)=\tanh (ux)\sim ux-\frac{u^{3}x^{3}}{3}+O(x^{5}),
\end{equation}
 where \( u,r \) are parameters. The variational method is based
on trial functions depending on two parameters, \( u \) and \( r \).
We can find the \( E \) total energy dependence on the \( u,r \)
parameters, for the sphaleron, (see the Fig.1). 

We then get the minimum energy for \( u_{0} \), and \( r_{0} \).
The trial functions for these parameters (\( u_{0} \), \( r_{0} \))
give us initial data for the numerical solutions. So, we have the
trial functions (solid lines in the Fig.1) and the numerical solutions
(dashed lines) from the initial conditions obtained. We use the Runge-Kutta
method (the shooting method \cite{num}) for our numerical calculation.
The trial functions are used to find the initial data close to zero,
if \( x \) goes to zero (\( x_{0} \)), and far from zero, if x goes
to infinity (\( x_{\infty } \)). We start from \( x_{0} \) and \( x_{\infty } \),
and get two solutions. Then we glue them together. If the numerical
method is correct then both of the numerical solutions should cover
each other. We obtain such results for the gauge field, (see the Fig.3)
and for the Higgs field, (see the Fig.4). 

It is also very interesting to find the sphaleron mass dependence
on the Higgs mass, (see the Fig.2). The rest mass of a sphaleron (without
dilatons) is rather large (for our parameters): \[
(M_{sph}\sim E=10.006\, \, TeV).\]
 The sphaleron remains a bubble of the old high temperature phase.
We construct also a trial function for the dilaton field: \begin{equation}
s(x)=s_{\infty }+ae^{-wx^{2}},
\end{equation}
 where \( w \) and \( a \) are parameters. We solve the set of the
differential equations (35 -37) using the trial functions (44 - 46),
which provide the initial conditions for a numerical estimate. The
method of calculation is the same as for the sphaleron without dilatons.
For the dilatonic sphaleron, we can also find the \( E \) total energy
dependence on the \( a \) parameter, (see the Fig.5). 

The minimum energy we get for \( a_{0}=-0.384 \) (with the same parameters
as before). The mass of the dilatonic sphaleron is decreased, as follows:
\begin{equation}
M_{dil.sph}\sim E_{d}=7.917\, \, \, \, TeV.
\end{equation}
 The trial functions at these parameters give us the initial conditions
for the numerical solutions. So, we have the trial functions (solid
lines), and the numerical solutions (dashed lines) in the Fig. 5.
The trial functions are used to find the initial data close to zero,
and far from zero. We get two solutions, again. Then we glue them
together. Both numerical solutions cover each other. We present the
cases for the gauge field, (see the Fig.7) and for the Higgs field,
(see the Fig.8). 

With respect to the dilatation symmetry of the dilaton field we can
shift the dilaton field and require its vanishing in the infinity.
The numerical solution is stable for the dilatonic field all over
the range of \( x \), (see the dashed line in the Fig.9). The numerical
solutions of the coupled system of the differential equations are
close to our trial functions. 

We have also found another local minimum for \( a>0 \), (see the
Fig.6). We see that it is very close to the sphaleron configuration
without dilatons. These solutions differ from one another in the dilatonic
cloud. 

In the first solution (if \( a_{0}=-0.384<0 \)) the Higgs field is
amplified (the dilaton field causes the increase of the energy of
the Higgs field) and the gauge field is suppressed (the dilaton field
reduces the energy of the gauge field) in the sphaleron. In the second
solution (if \( a>0 \)) we have the reverse.\\

\section{CONCLUSIONS}

Numerical solutions suggest that the sphaleron possesses an 'onion-like'
structure. In the small inner core, the scalar field decreases, with
global gauge symmetry restoration \( SU(2)\times U(1) \). In the
middle layer, the gauge field undergoes sudden change. It is very
interesting that the sphaleron coupled to the dilaton field has also
an outer shell, where the dilaton field changes drastically. Our solutions
describe the shapes of Higgs field and gauge field inside the sphaleron,
as well as the change of the dilaton cloud surrounding the sphaleron.
Such a cloud is large and extends far outside the sphaleron \cite{km:1996}.
The sphalerons might be created during the first order phase transition
in the expanding universe. The bubbles left after the phase transition
could break the CP and C symmetry on their walls and could cause the
breaking of the baryonic symmetry. Further increase of the energy
of the dilaton fields causes the increase of the energy (mass) of
the sphaleron. However, a decrease of the energy of the sphaleron
causes the increase of the tunneling probability to the new nontrivial
configuration. This effect generates the increase of the probability
of baryogenesis \cite{pe}. With the sphalerons, the gauge field and
the Higgs field configurations are very similar to the configurations
of the same fields without dilatons. Comparing the three Figs. 7,
8, 9 we can conclude that the dilatonic area is much bigger than the
area of the change of the gauge field and the Higgs field. The conclusion
drawn is that the dilaton field in general causes the increase of
the mass of the sphaleron. However, there exists a minimal configuration
at which the energy of the sphaleron is decreased. A sphaleron is
said to be concentric with the following configuration: the Higgs
field and the gauge field are embedded in the dilaton cloud; close
to the center there is a change of the Higgs field, and then a change
of the gauge field, both of them penetrated by a change of the dilaton
field \cite{km:1996}. Now, we can see that dilatons have a small
influence the Higgs field and the gauge field configurations. That
means that a topological 'hedgehog' structure of the sphaleron is
not modified. The sphaleron properties depend strongly on the dilaton
field which appears from more dimensional spacetime compactification. 

However, in the theories inspired by the Kaluza-Klein theory with
large extra dimensions, new interactions with massive (\( \sim M \))
Kaluza-Klein gravitons also take place. Due to the topological charge,
the sphaleron solutions are stable in the four-dimensional spacetime.
The interactions with Kaluza-Klein gravitons \( h_{\mu \nu }^{{\textbf {n}}}(x) \)
may cause the disintegration of the sphaleron. 

There are also interesting spherically symmetric dilaton solutions
coupled to the gauge field in the monopole case \cite{kk:1999} and
they may influence the monopole catalysis of baryogenesis.

\section*{FIGURES}

\newpage

\begin{figure}
\par\centering \resizebox*{10cm}{!}{\includegraphics{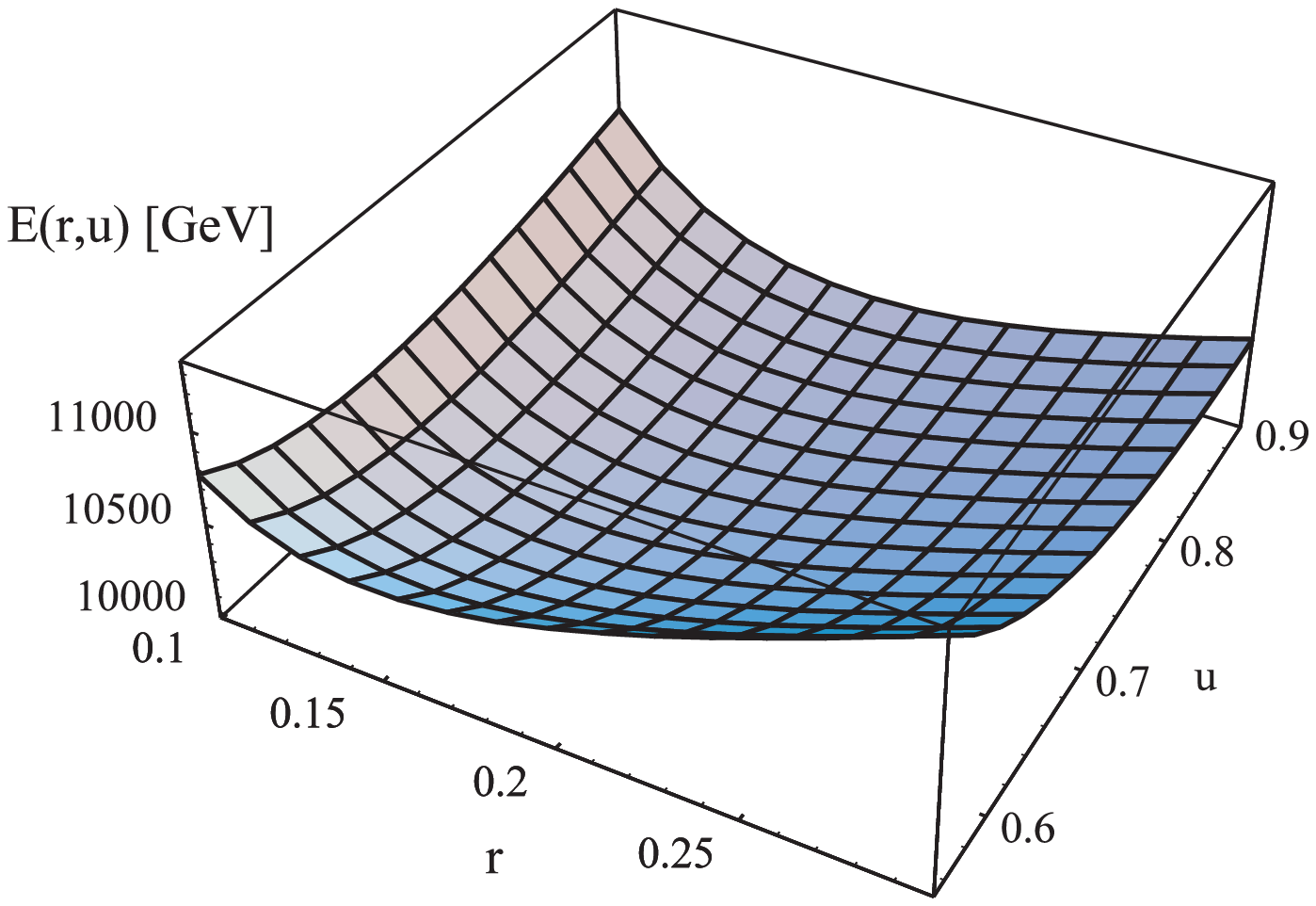} } \par{}

\caption{The total energy of the sphaleron in the standard model at the parameters
(\protect\protect\protect\protect\protect\( \varepsilon =1.152\protect \protect \protect \protect \protect \),
\protect\protect\protect\protect\protect\( M_{H}=120\protect \protect \protect \protect \protect \)
GeV); the minimum energy (E(r,u) = 10 006 GeV) is for the values:
\protect\protect\protect\protect\protect\( r=0.1735\protect \protect \protect \protect \protect \),
\protect\protect\protect\protect\protect\( u=0.7586\protect \protect \protect \protect \protect \).}\end{figure}

\begin{figure}
\par\centering \resizebox*{10cm}{!}{\includegraphics{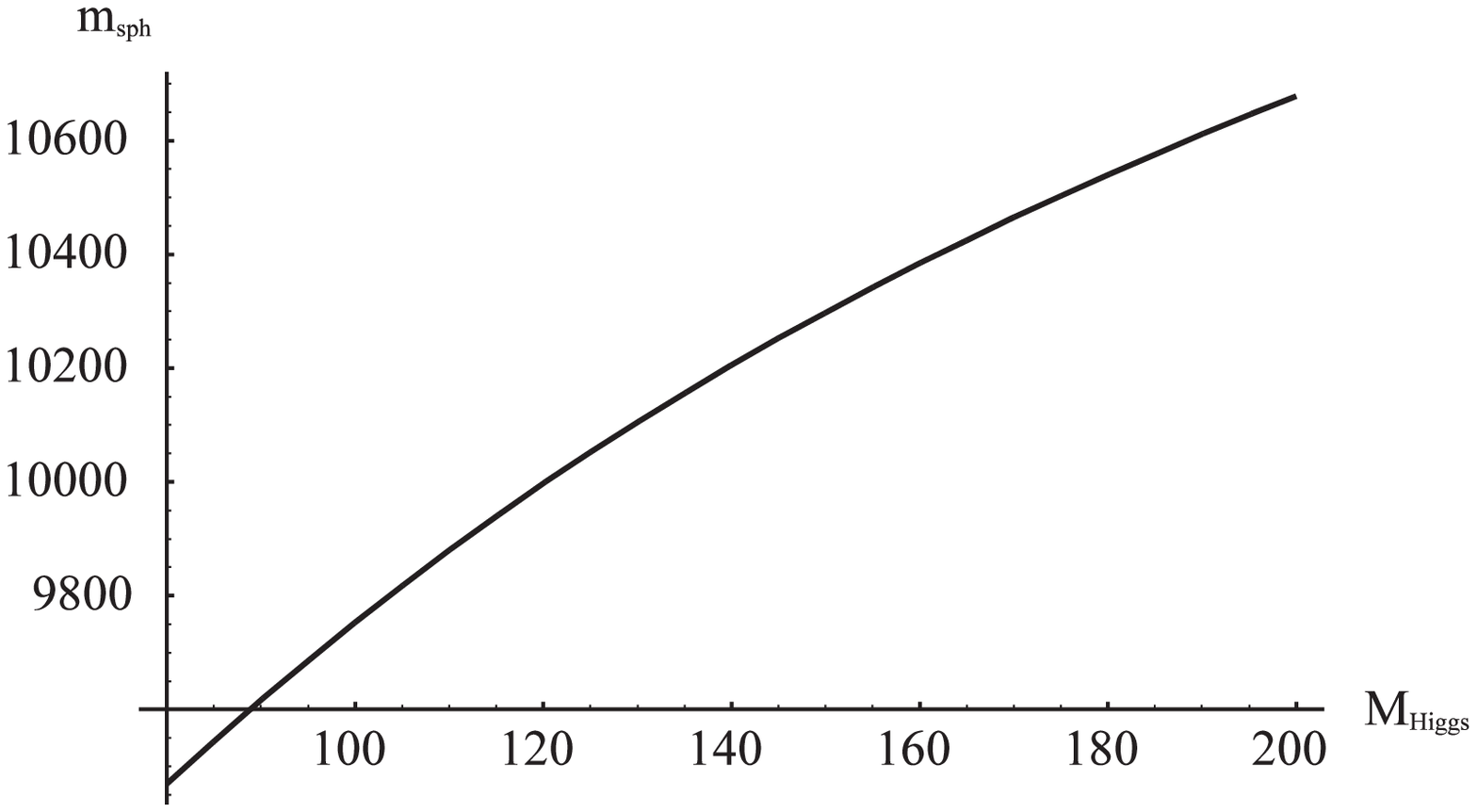} } \par{}

\caption{The dependence of the mass of the sphaleron on the Higgs mass (\protect\protect\protect\protect\protect\( M_{H}\protect \protect \protect \protect \protect \)).}\end{figure}

\begin{figure}
\par\centering \resizebox*{10cm}{!}{\includegraphics{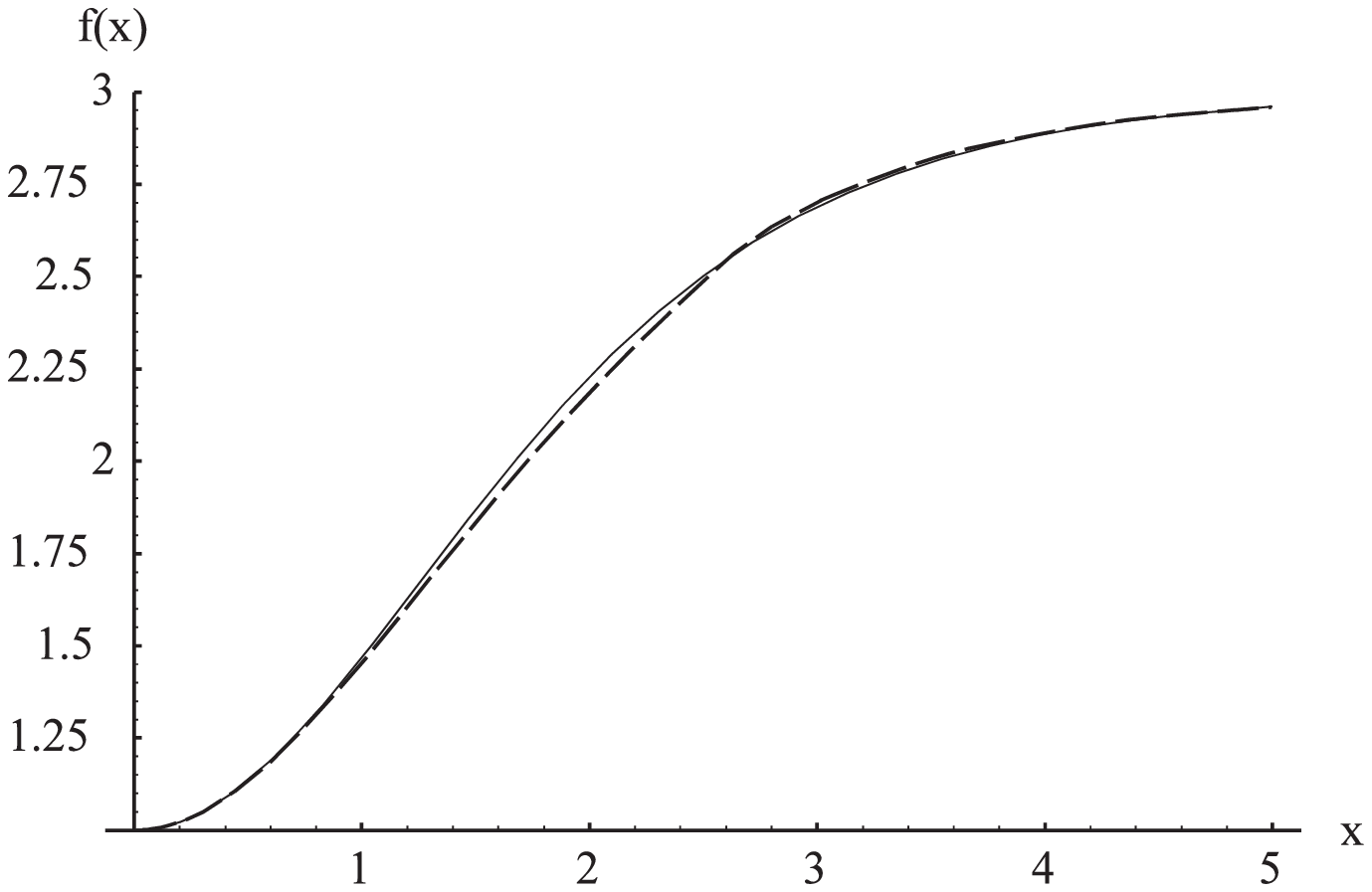} } \par{}

\caption{The dependence of the gauge field of the sphaleron on x, at the parameters
\protect\protect\protect\protect\protect\( r=0.2777\protect \protect \protect \protect \protect \),
\protect\protect\protect\protect\protect\( u=0.8193\protect \protect \protect \protect \protect \).}\end{figure}

\begin{figure}
\par\centering \resizebox*{10cm}{!}{\includegraphics{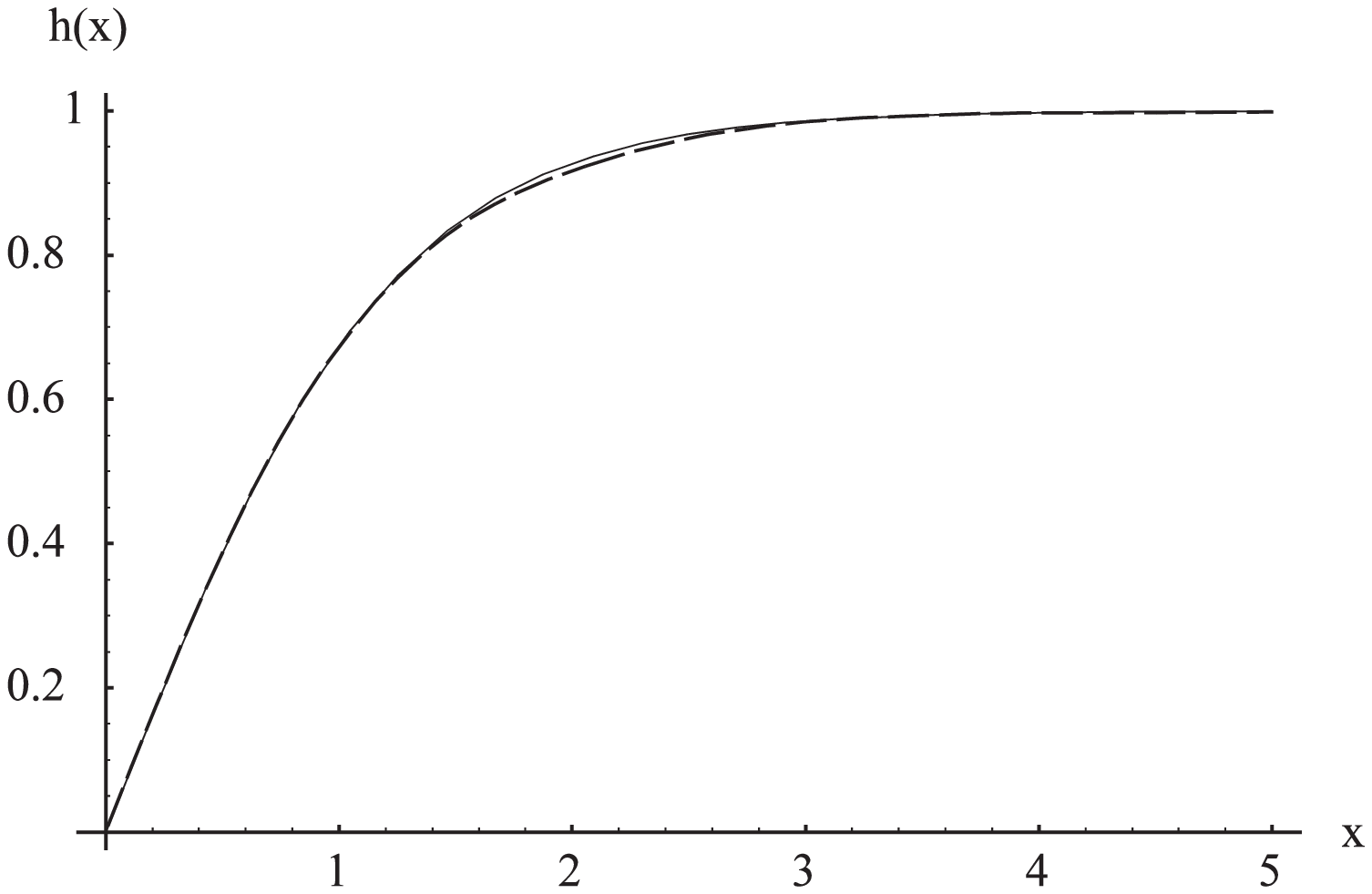} } \par{}

\caption{The dependence of the Higgs field of the sphaleron on x, at the parameters
\protect\protect\protect\protect\protect\( r=0.2777\protect \protect \protect \protect \protect \),
\protect\protect\protect\protect\protect\( u=0.8193\protect \protect \protect \protect \protect \).}\end{figure}

\begin{figure}
\par\centering \resizebox*{10cm}{!}{\includegraphics{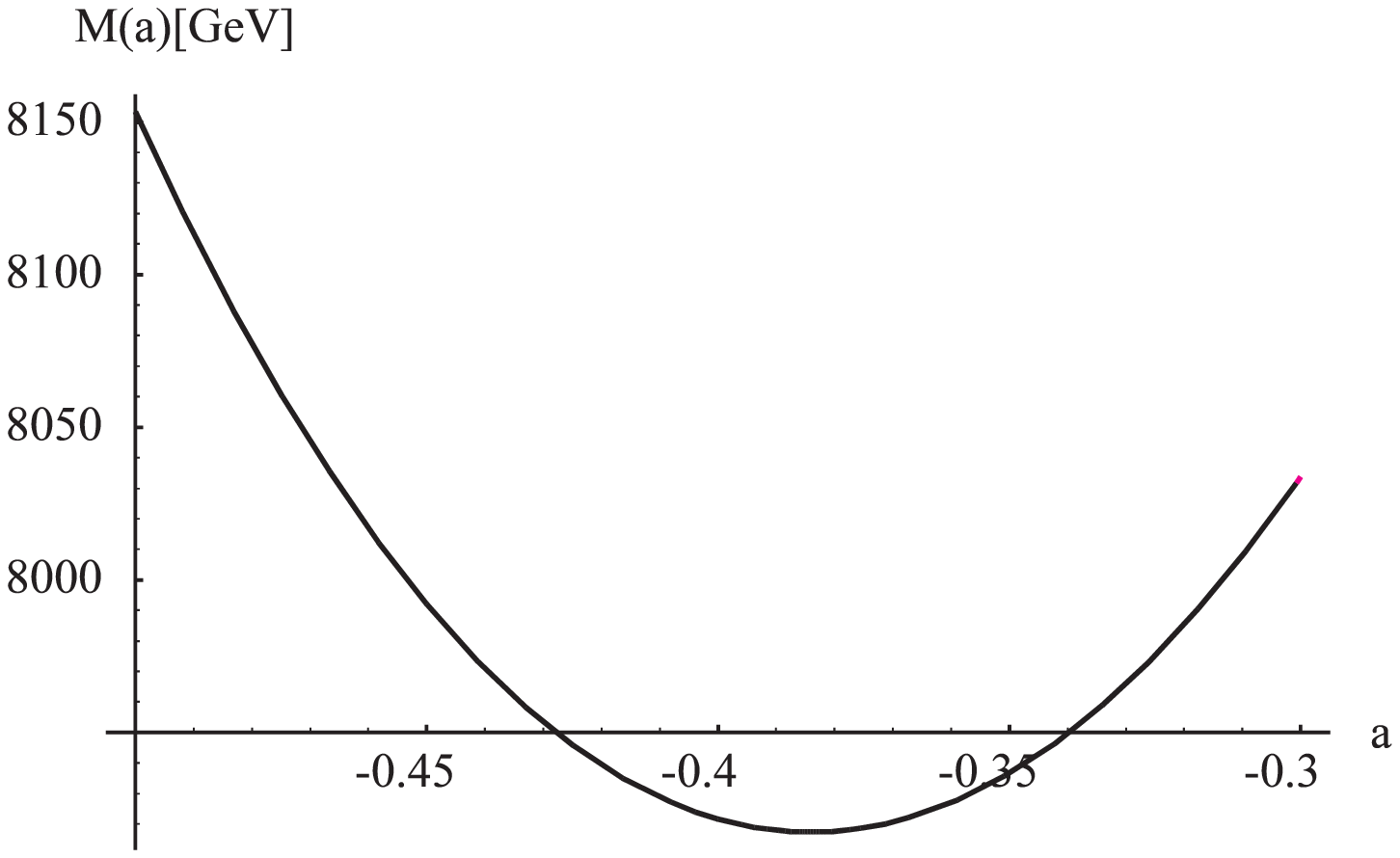} } \par{}

\caption{The dependence of the mass of the dilatonic sphaleron on \textit{a}
parameter (\protect\protect\protect\protect\protect\( a<0\protect \protect \protect \protect \protect \)),
at the parameters: \protect\protect\protect\protect\protect\( r=0.2958\protect \protect \protect \protect \protect \),
\protect\protect\protect\protect\protect\( u=0.9862\protect \protect \protect \protect \protect \),
\protect\protect\protect\protect\protect\( w=2.47\, \, 10^{-8}\protect \protect \protect \protect \protect \).
The minimum is for: \protect\protect\protect\protect\protect\( a=-0.3540\protect \protect \protect \protect \protect \),
\protect\protect\protect\protect\protect\( M(a)=7917.36\protect \protect \protect \protect \protect \)
GeV.}\end{figure}

\begin{figure}
\par\centering \resizebox*{10cm}{!}{\includegraphics{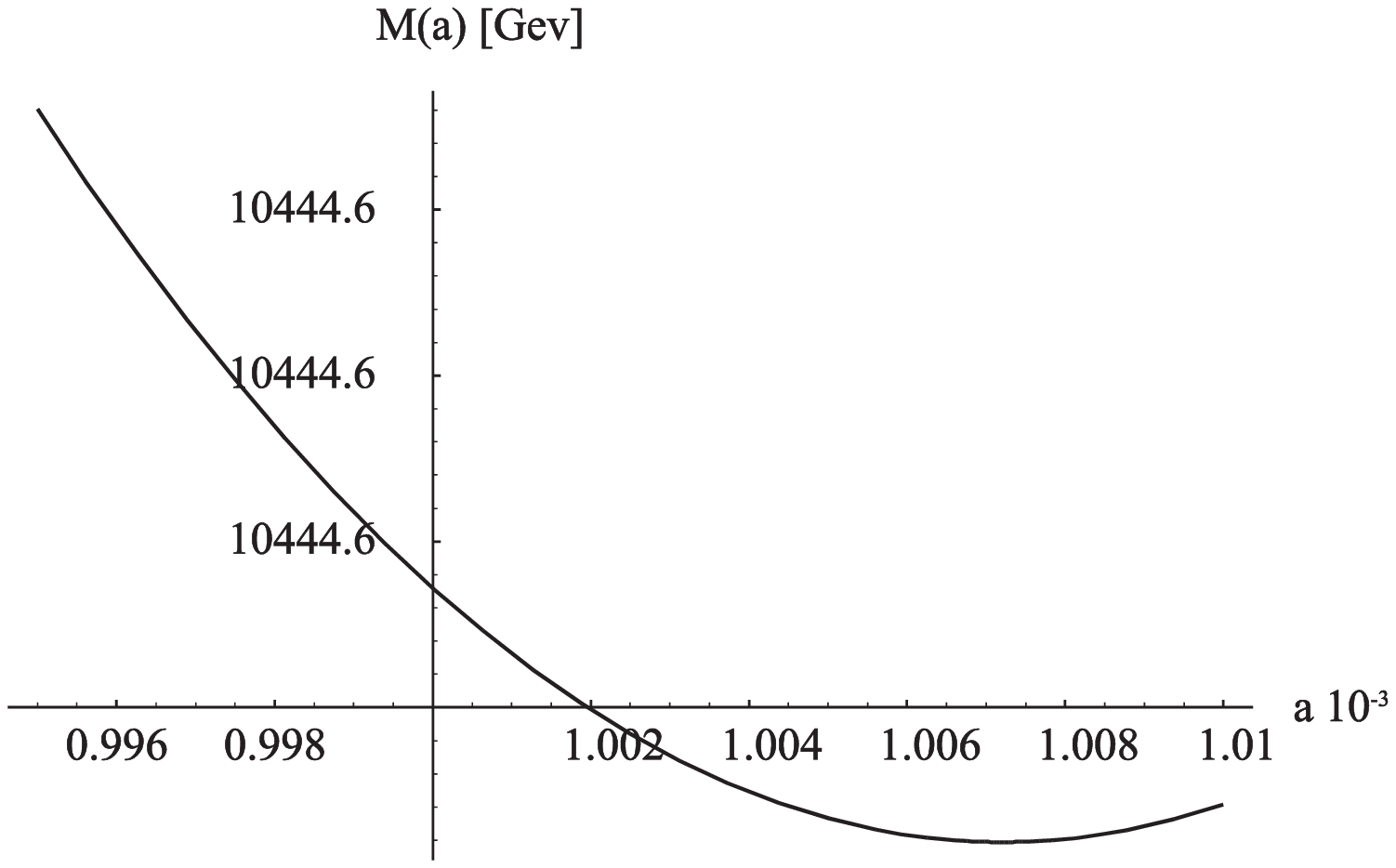} } \par{}

\caption{The dependence of the mass of the dilatonic sphaleron on \textit{a}
parameter (\protect\protect\protect\protect\protect\( a>0\protect \protect \protect \protect \protect \)),
at the parameters: \protect\protect\protect\protect\protect\( r=0.2958\protect \protect \protect \protect \protect \),
\protect\protect\protect\protect\protect\( u=0.9862\protect \protect \protect \protect \protect \),
\protect\protect\protect\protect\protect\( w=2.47\, \, 10^{-8}\protect \protect \protect \protect \protect \).}\end{figure}

\begin{figure}
\par\centering \resizebox*{10cm}{!}{\includegraphics{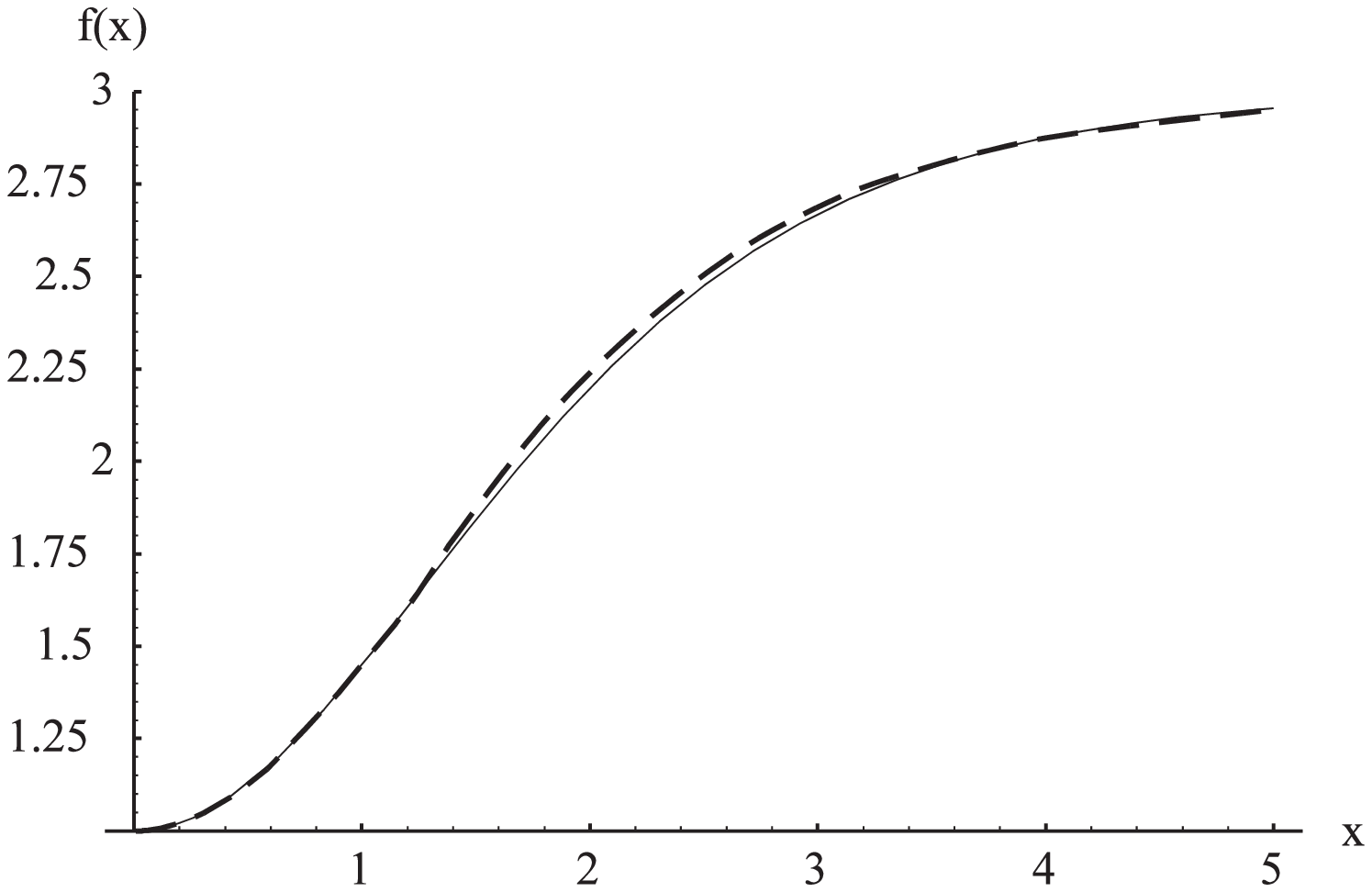} } \par{}

\caption{The dependence of the gauge field of the dilatonic sphaleron on x,
at the parameters \protect\protect\protect\protect\protect\( r=0.2950\protect \protect \protect \protect \protect \),
\protect\protect\protect\protect\protect\( u=0.9862\protect \protect \protect \protect \protect \),
\protect\protect\protect\protect\protect\( a=-0.3540\protect \protect \protect \protect \protect \),
\protect\protect\protect\protect\protect\( w=2.47\, \, 10^{-8}\protect \protect \protect \protect \protect \).}\end{figure}

\begin{figure}
\par\centering \resizebox*{10cm}{!}{\includegraphics{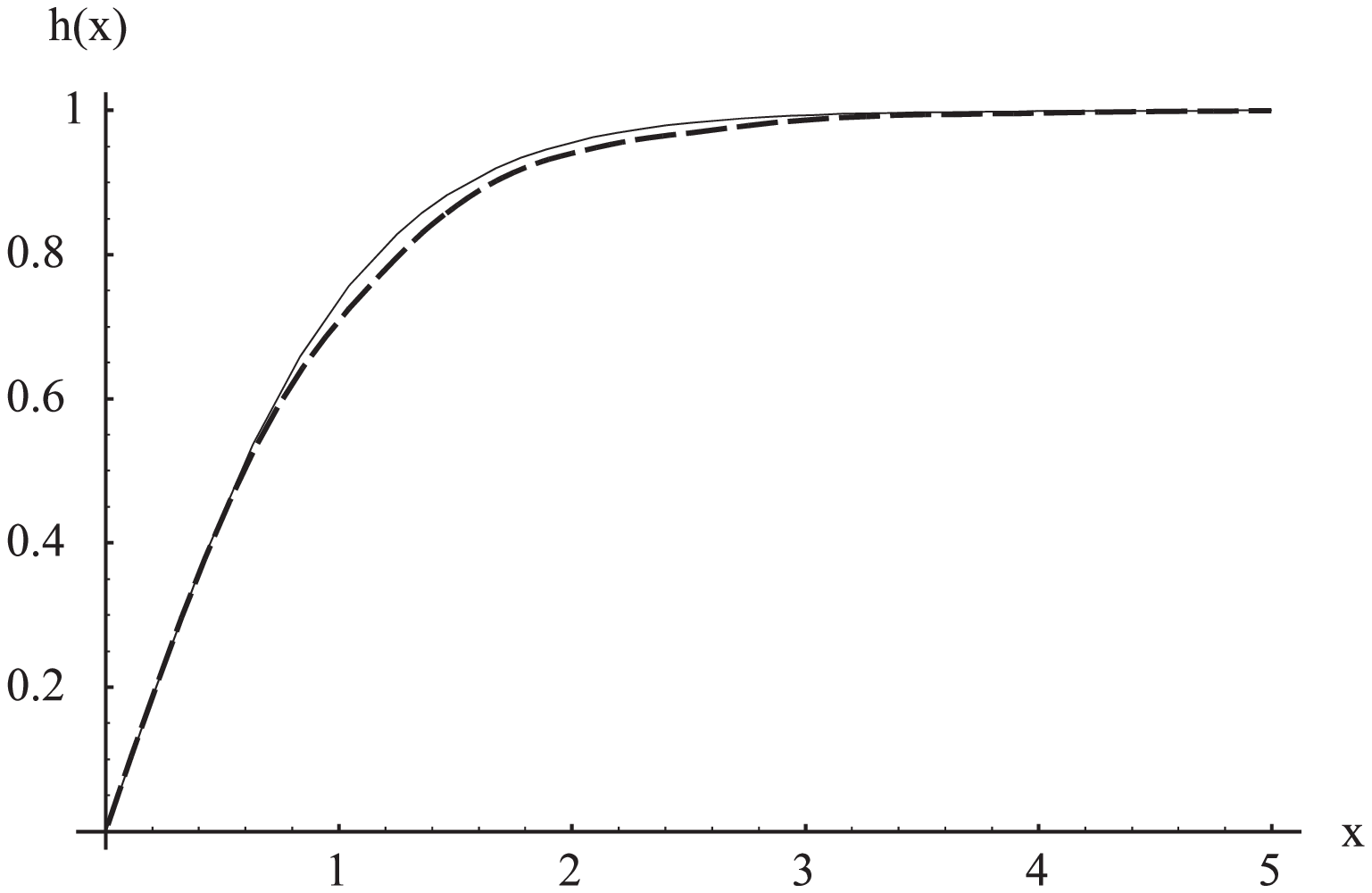} } \par{}

\caption{The dependence of the Higgs field of the dilatonic sphaleron on x,
at the parameters \protect\protect\protect\protect\protect\( r=0.2950\protect \protect \protect \protect \protect \),
\protect\protect\protect\protect\protect\( u=0.9862\protect \protect \protect \protect \protect \),
\protect\protect\protect\protect\protect\( a=-0.3540\protect \protect \protect \protect \protect \),
\protect\protect\protect\protect\protect\( w=2.47\, \, 10^{-8}\protect \protect \protect \protect \protect \).}\end{figure}

\begin{figure}
\par\centering \resizebox*{10cm}{!}{\includegraphics{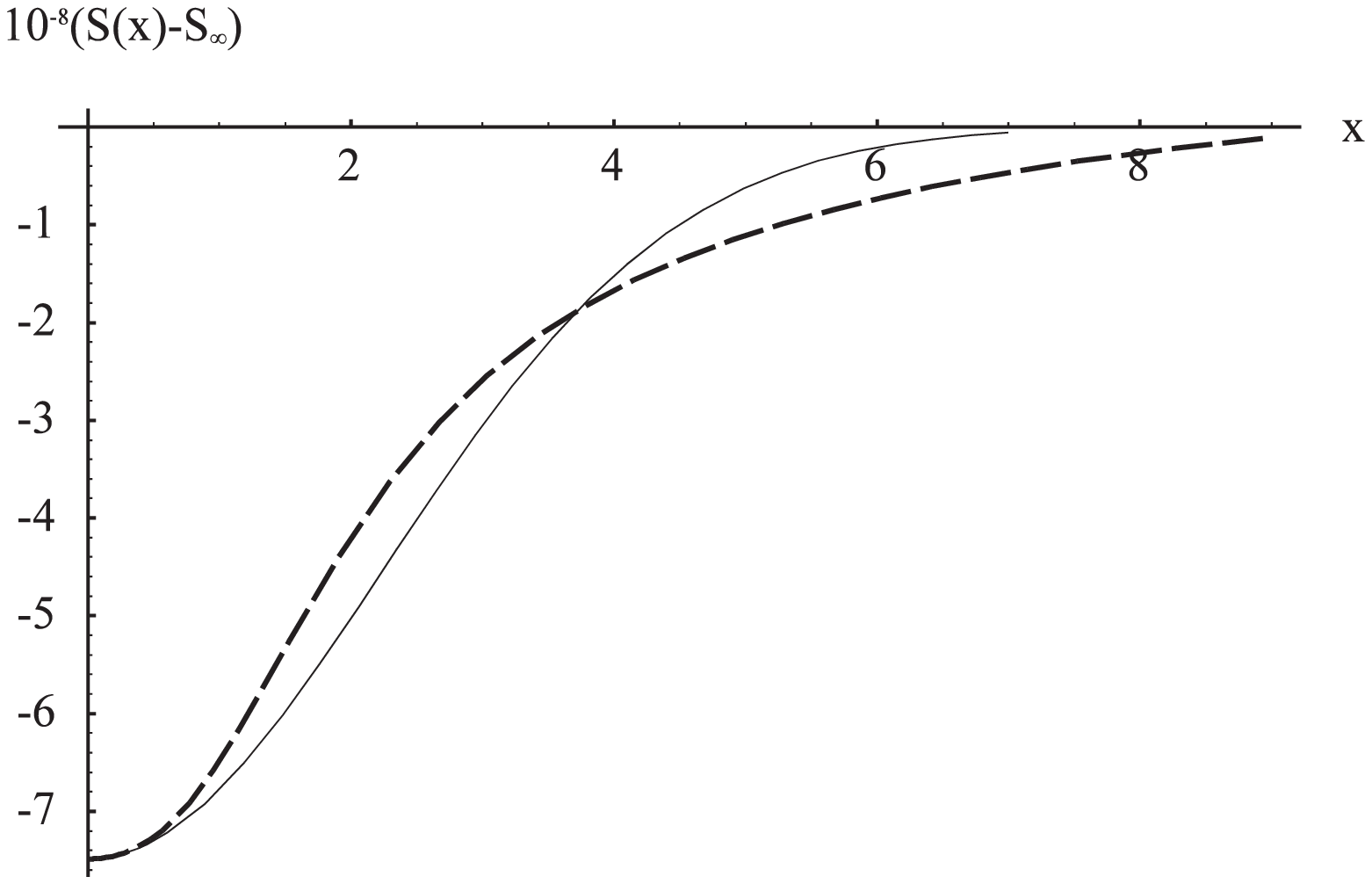} } \par{}

\caption{The dependence of the dilaton field on x, at the parameters \protect\protect\protect\protect\protect\( r=0.2777\protect \protect \protect \protect \protect \),
\protect\protect\protect\protect\protect\( u=0.9862\protect \protect \protect \protect \protect \),
\protect\protect\protect\protect\protect\( a=-0.3540\protect \protect \protect \protect \protect \),
\protect\protect\protect\protect\protect\( w=2.47\, \, 10^{-8}\protect \protect \protect \protect \protect \).}\end{figure}

\newpage

\section*{Figure captions}

Figure 1. The total energy of the sphaleron in the standard model
at the parameters (\protect\( \varepsilon =1.152\protect  \), \protect\( M_{H}=120\protect  \)
GeV); the minimum energy (E(r,u) = 10 006 GeV) is for the values:
\protect\( r=0.1735\protect  \), \protect\( u=0.7586\protect  \).\\
 Figure 2. The dependence of the mass of the sphaleron on the Higgs
mass (\protect\( M_{H}\protect  \)). \\
 Figure 3. The dependence of the gauge field of the dilatonic sphaleron
on x, at the parameters \protect\( r=0.2950\protect  \), \protect\( u=0.9862\protect  \),
\protect\( a=-0.3540\protect  \), \protect\( w=2.47\, \, 10^{-8}\protect  \).\\
 Figure 4. The dependence of the Higgs field of the dilatonic sphaleron
on x, at the parameters \protect\( r=0.2950\protect  \), \protect\( u=0.9862\protect  \),
\protect\( a=-0.3540\protect  \), \protect\( w=2.47\, \, 10^{-8}\protect  \)\\
 Figure 5. The dependence of the mass of the dilatonic sphaleron on
\textit{a} parameter (\protect\( a<0\protect  \)), at the parameters:
\protect\( r=0.2958\protect  \), \protect\( u=0.9862\protect  \),
\protect\( w=2.47\, \, 10^{-8}\protect  \). The minimum is for: \protect\( a=-0.3540\protect  \),
\protect\( M(a)=7917.36\protect  \) GeV.\\
 Figure 6. The dependence of the mass of the dilatonic sphaleron on
\textit{a} parameter (\protect\( a>0\protect  \)), at the parameters:
\protect\( r=0.2958\protect  \), \protect\( u=0.9862\protect  \),
\protect\( w=2.47\, \, 10^{-8}\protect  \)\\
 Figure 7. The dependence of the gauge field of the dilatonic sphaleron
on x, at the parameters \protect\( r=0.2950\protect  \), \protect\( u=0.9862\protect  \),
\protect\( a=-0.3540\protect  \), \protect\( w=2.47\, \, 10^{-8}\protect  \).\\
 Figure 8. The dependence of the Higgs field of the dilatonic sphaleron
on x, at the parameters \protect\( r=0.2950\protect  \), \protect\( u=0.9862\protect  \),
\protect\( a=-0.3540\protect  \), \protect\( w=2.47\, \, 10^{-8}\protect  \).\\
 Figure 9. The dependence of the dilaton field on x, at the parameters
\protect\( r=0.2777\protect  \), \protect\( u=0.9862\protect  \),
\protect\( a=-0.3540\protect  \), \protect\( w=2.47\, \, 10^{-8}\protect  \).\\

\end{document}